\documentclass[twocolumn,trackchanges]{aastex701}

\pdfoutput=1
\usepackage{lineno}
\usepackage{amsmath}        
\usepackage{placeins}
\usepackage{graphicx}
\usepackage{booktabs}       
\usepackage{array}          
\usepackage{multirow}       
\usepackage{graphicx}       
\usepackage{dcolumn}        
\usepackage{xspace}         
\usepackage{amsmath}
\usepackage{amssymb}
\usepackage[american]{babel}
\usepackage{booktabs}
\usepackage{dcolumn}  
\usepackage[T1]{fontenc}  
\usepackage[]{hyperref}  
\usepackage[utf8]{inputenc}  
\usepackage{multirow}
\usepackage{txfonts}  
\usepackage{url}
\usepackage[dvipsnames]{xcolor}

\usepackage{float}
\usepackage{tabularx}
\usepackage{booktabs} 
\usepackage{ragged2e}      
\renewcommand{\arraystretch}{1.5} 

\DeclareUnicodeCharacter{2212}{-}
\makeatletter

\newcommand{\Rmnum}[1]{\expandafter\@slowromancap\romannumeral #1@}
\definecolor{greenW}{rgb}{0.0, 0.55, 0.1}

\usepackage[normalem]{ulem}

\makeatother

\renewcommand {\arraystretch}{1.3}

\hypersetup{
	citecolor = red,
	linkcolor = blue,
	urlcolor = magenta
}

\submitjournal{ApJL}

\begin{document}
	
\title{Sound-Horizon-Agnostic Inference of the Hubble Constant and Neutrino Mass from BAO, CMB Lensing, and Galaxy Weak Lensing and Clustering}

\author{Helena Garc\'ia Escudero}
\thanks{E-mail: \url{garciaeh@uci.edu}}
\affiliation{Center for Cosmology, Department of Physics and Astronomy, University of California, Irvine, California 92697-4575, USA}
\email{garciaeh@uci.edu}

\author{Seyed Hamidreza Mirpoorian}
\thanks{E-mail: \url{smirpoor@sfu.ca}}
\affiliation{Department of Physics, Simon Fraser University, Burnaby, British Columbia, Canada V5A 1S6}
\email{smirpoor@sfu.ca}

\author{Levon Pogosian}
\thanks{E-mail: \url{levon@sfu.ca}}
\affiliation{Department of Physics, Simon Fraser University, Burnaby, British Columbia, Canada V5A 1S6}
\email{levon@sfu.ca}

\begingroup
\renewcommand\thefootnote{}
\footnote{The authors contributed equally to this work and are listed alphabetically \\ by last name.}
\endgroup

\begin{abstract}

We present a sound-horizon-agnostic determination of the Hubble constant, $H_0$, by combining DESI DR2 baryon acoustic oscillation (BAO) data with the latest cosmic microwave background (CMB) lensing measurements from Planck, ACT, and SPT-3G, the angular size of the CMB acoustic scale, Dark Energy Survey Year-3 ($3\times2$-pt) galaxy weak lensing and clustering correlations, and the Pantheon+ supernova sample. In this analysis, The sound horizon at the drag epoch, $r_d$, is treated as a free parameter. By combining uncalibrated comoving distances from BAO and supernovae with constraints on the matter density $\Omega_m h^2$ from CMB and galaxy lensing/clustering, we break the $r_d$–$H_0$ degeneracy and obtain $H_0 = 70.0 \pm 1.7$ km/s/Mpc when the sum of the neutrino masses is fixed at $\Sigma m_\nu = 0.06$ eV. With an informative prior on the amplitude of primordial fluctuations, $A_s$, we find $H_0 = 70.03 \pm 0.97$ km/s/Mpc. Allowing $\Sigma m_\nu$ to vary, we find that the neutrino mass is weakly constrained and strongly prior-dependent. Consequently, the inferred $H_0$ is sensitive to the choice of the $\Sigma m_\nu$ prior, with a uniform prior biasing results toward larger neutrino masses and higher $H_0$, while a logarithmic prior reduces this bias significantly. Forecasts for the completed DESI BAO program, combined with Simons-Observatory-like CMB lensing, next-generation $3\times2$-pt data, and expanded supernova samples predict $\sigma(H_0) \simeq 0.67$ km/s/Mpc with fixed $\Sigma m_\nu$, and $\sigma(H_0) \simeq 1.1$ km/s/Mpc with $\Sigma m_\nu < 0.133$ ($<0.263$) eV at 68\% (95\%) CL when $\Sigma m_\nu$ is varied.

\end{abstract}

\section{Introduction}
\label{sec:intro}

Determining the Hubble constant, $H_0$, with high precision is both a fundamental goal and an ongoing challenge in modern cosmology. Despite the success of the standard $\Lambda$CDM model in explaining a wide range of cosmological observations, a significant tension persists between the value $H_0 = 67.36 \pm 0.54$ km/s/Mpc inferred from \textit{Planck} measurements of the cosmic microwave background (CMB), under the assumption of $\Lambda$CDM~\citep{Planck:2018vyg}, and the value $H_0 = 73.04 \pm 1.04$ km/s/Mpc obtained from Cepheid-based distance ladder measurements~\citep{Riess:2021jrx}.

CMB analyses rely on the well-established physics of recombination to determine the sound horizon at photon--baryon decoupling, which directly impacts the extracted cosmological parameters. We distinguish between the comoving sound horizon at photon decoupling, $r_\star$, which sets the physical scale of the acoustic features observed in the CMB anisotropies, and the drag-epoch sound horizon, $r_d$, which governs the characteristic BAO standard ruler measured in large-scale structure. The two distances correspond to closely related epochs and are tightly correlated. The Hubble tension may indicate the need for additional physics during the epoch of recombination, such as primordial magnetic fields~\citep{Jedamzik:2013gua,Jedamzik:2018itu,Jedamzik:2020krr,Jedamzik:2025cax}, early dark energy~\citep{Karwal:2016vyq,Poulin:2018cxd,Smith:2019ihp}, extra relativistic species~\citep{Bernal:2016gxb,Escudero:2024uea}, neutrino self-interactions~\citep{Kreisch:2019yzn,Pan:2019gop}, or variations of fundamental constants that alter recombination physics~\citep{Hart:2019dxi,Hart:2021kad,Baryakhtar:2024rky}, among other proposed mechanisms~\citep{DiValentino:2021izs,Schoneberg:2021qvd,Escudero:2022rbq}. These scenarios would modify the standard determination of $r_d$, making it smaller and thereby increasing the CMB-inferred value of $H_0$. Notably, the $\Lambda$CDM relation between the two sound horizons is unchanged in these models, with $r_d \approx 1.019 r_\star$.

This motivates the exploration of methods for constraining $H_0$ that do not rely on modeling $r_d$. Several recent studies~\citep{Philcox:2020xbv,Baxter:2020qlr,Farren:2021grl,Philcox:2022sgj,Zaborowski:2024car,SPT-3G:2025zuh} have utilized datasets that probe the scale of matter--radiation equality, $k_{\rm eq}$, and hence the physical matter density $\Omega_m h^2$, in combination with uncalibrated supernova (SN) luminosity distances, which constrain the present-day matter fraction $\Omega_m$, to obtain a sound-horizon-free measurement of $H_0$. Both the full-shape galaxy power spectrum $P(k)$ and the CMB lensing convergence spectrum $C_\ell^{\kappa\kappa}$ are sensitive to $k_{\rm eq}$. In full-shape analyses, $P(k)$ is decomposed into its smooth (the component that constrains $k_{\rm eq}$) and oscillatory (BAO) parts, and $r_d$ is marginalized over in the oscillatory part to eliminate dependence on the sound horizon. While this method preserves some of the calibration-independent angular diameter distance information encoded in the BAO wiggles, it does not fully exploit the constraining power of the complete set of transverse, radial, and volume-averaged BAO observables.

In this work, we employ an alternative approach introduced in~\citep{Pogosian:2020ded}, combining uncalibrated transverse, radial, and volume-averaged BAO measurements, which constrain $r_d H_0$ and $\Omega_m$, with CMB and galaxy lensing, which provide a constraint on $\Omega_m h^2$ and break the $r_d$--$H_0$ degeneracy. Here, $r_d$ is treated as a free primary parameter and is measured from the data alongside $H_0$. We use BAO measurements from the DESI DR2 release~\citep{DESI:2025zgx}, the combination~\citep{SPT-3G:2025zuh} of CMB lensing convergence spectra $C_\ell^{\kappa\kappa}$ from \textit{Planck}~\citep{Planck:2020olo}, the Atacama Cosmology Telescope (ACT DR6)~\citep{ACT:2023dou,ACT:2023ubw,ACT:2023kun}, and the South Pole Telescope (SPT-3G)~\citep{SPT-3G:2024atg}, hereafter referred to as APS, along with the DES Year-3 three two-point galaxy weak lensing and clustering correlation functions ($3\times2$-pt, DESY3)~\citep{DES:2021wwk,DES:2025xii}. In addition, we include the CMB acoustic scale angle, $\theta_{\star}$, treating it as an additional transverse BAO point at redshift $z = 1090$. We also include the Pantheon+ compilation of uncalibrated SN magnitudes~\citep{Brout:2022vxf}, which add an additional constraint on $\Omega_m$; however, this contribution is not significant since the uncalibrated BAO data alone already constrain $\Omega_m$ well.

In addition to providing an early-Universe-independent constraint on $H_0$, our approach also enables, in principle, a measurement of the sum of neutrino masses, $\Sigma m_\nu$, through the sensitivity of CMB lensing and galaxy lensing and clustering to the suppression of small-scale structure, especially when combined with a weak external prior on the amplitude of primordial fluctuations $A_s$. By combining DESI DR2 BAO and $\theta_{\star}$ with CMB lensing from APS, DESY3 galaxy weak lensing and clustering, and the Pantheon+ SN, we obtain $H_0 = 70.0 \pm 1.7$ km/s/Mpc when $\Sigma m_\nu$ is fixed at the minimal value of $0.06$ eV. With an additional informative prior on $A_s$, we obtain $H_0 = 70.03 \pm 0.97$ km/s/Mpc and $r_d = 144.8 \pm 1.6$~Mpc. Allowing $\Sigma m_\nu$ to vary, we find $H_0 = 75.3^{+3.3}_{-4.0}$ km/s/Mpc and $\Sigma m_\nu < 1.11$ eV (95\% CL) without an informative prior on $A_s$, and $H_0 = 73.9 \pm 2.2$ km/s/Mpc with $\Sigma m_\nu = 0.46^{+0.40}_{-0.45}$ eV (95\% CL) when the prior on $A_s$ is included. We also perform forecasts showing that future CMB lensing data from a Simons-Observatory-like (SO-L) experiment~\citep{SimonsObservatory:2025wwn} and future galaxy weak lensing and clustering measurements, when combined with data from the completed DESI BAO program and an expanded SN dataset, can constrain $H_0$ with a precision of $\sim 0.67$ km/s/Mpc if $\Sigma m_\nu$ is fixed, and $\sim 1.1$ km/s/Mpc when $\Sigma m_\nu$ is varied, with $\Sigma m_\nu < 0.26$ eV (95\% CL). 

The structure of this paper is as follows: Section~\ref{sec:rdCAMB} introduces our approach to using the BAO observables in an $r_d$-independent way and discusses the sensitivity of the datasets used in this analysis to the parameters of interest. Section~\ref{sec:lik} presents the details of the analysis, including the data likelihoods, parameter prior assumptions, and the forecast methodology. In Section~\ref{sec:results}, we present the constraints derived from current data and the forecasts, with and without varying $\Sigma m_\nu$. We summarize our findings in Section~\ref{sec:concl}.

\section{Constraining cosmology with the sound horizon as a free parameter}
\label{sec:rdCAMB}

\subsection{BAO Observables}

BAO provide standard-ruler distance measurements through three types of observables~\citep{SDSS:2005xqv}: the transverse BAO scale, the line-of-sight BAO scale, and the isotropic average of the two. The transverse observable measures the angular size of the sound horizon $r_d$ and is defined as
\begin{equation}
    \beta_{\perp}(z) = \frac{D_M(z)}{r_d},
\end{equation}
where \( D_M(z) \) is the comoving distance to redshift $z$. For illustration, we focus on \( \beta_{\perp}(z) \), though the methodology described below applies equally to the radial and isotropic BAO observables.

Assuming a spatially flat \(\Lambda\)CDM model and neglecting radiation (an acceptable approximation at the relevant redshifts), \( \beta_{\perp}(z) \) can be written as
\begin{align}
    \beta_{\perp}(z) 
    &= \frac{1}{r_d} \int_0^z \frac{c\,dz'}{H(z')} \nonumber \\
    &= \frac{2998\,\text{Mpc}}{r_d h} \int_0^z \frac{dz'}{\sqrt{\Omega_m(1+z')^3 + (1 - \Omega_m)}},
\end{align}
where \( c/H_0 = 2998\,\text{Mpc}/h \), and \( h = H_0 / (100\,\text{km/s/Mpc}) \). This shows that BAO observations at multiple redshifts constrain both the matter density fraction \( \Omega_m \) and the product \( r_d h \). Breaking the degeneracy between \( r_d \) and \( h \) requires additional information, such as the standard-model prediction for $r_d$ combined with a Big Bang Nucleosynthesis (BBN) prior on the baryon density \( \omega_b \)~\citep{Addison:2017fdm}. An alternative, explored in~\citep{Pogosian:2020ded,Pogosian:2024ykm} and adopted here, is to supplement BAO with a constraint on \( \Omega_m h^2 \) while treating $r_d$ as a free parameter, thereby avoiding assumptions about the sound horizon.

The additional constraint on \( \Omega_m h^2 \) can come from CMB lensing measurements, which are sensitive to the matter–radiation equality scale \( k_{\rm eq} \) and thus constrain the matter-to-radiation density ratio. As a consistency test, one may also adopt a Gaussian prior on \( \Omega_m h^2 \) from the Planck best-fit \(\Lambda\)CDM model and check whether the $H_0$ and $r_d$ values inferred from BAO agree with those in the best-fit model~\citep{Pogosian:2020ded,Pogosian:2024ykm}.

Unlike full-shape $P(k)$ approaches~\citep{Philcox:2020xbv,Philcox:2022sgj,Zaborowski:2024car}, this method makes use of all three BAO observables at each redshift, maximizing the geometrical information encoded in the data. In particular, it enables a constraint on $H_0$ that is less dependent on SN-derived values of \( \Omega_m \), since BAO alone already provide a strong constraint.

Finally, one can include the CMB acoustic scale angle, \( \theta_{\star} \), treating it as an additional transverse BAO data point at \( z = 1100 \), as discussed in Sec.~\ref{thetaCMB}.

\subsection{CMB Lensing, Galaxy Lensing and Clustering}

While the CMB lensing convergence spectrum \(C_\ell^{\kappa\kappa}\) depends the shape of the matter power spectrum, including the scale of matter--radiation equality, most of the sensitivity of \(C_\ell^{\kappa\kappa}\)to $\Omega_m h^2$ comes from the amplitude of the spectrum. As shown in~\citep{Planck:2015mym,Baxter:2020qlr}, CMB lensing constrains the combination \( A_s (\Omega_m h^2)^\alpha \), where \( \alpha > 0 \). Owing to projection effects, \( C_\ell^{\kappa\kappa} \) is nearly insensitive to the acoustic oscillation features in \( P(k) \), and hence to the value of \( r_d \), as we demonstrate below. However, CMB lensing does depend on the baryon fraction of the total matter density, since baryons cluster later than dark matter and their abundance affects the growth factor. For this reason, even though we treat \( r_d \) as a free parameter, we still apply the BBN prior on \( \omega_b \). 

\begin{figure}[htbp!]
    \centering
    \includegraphics[width=0.45\textwidth]{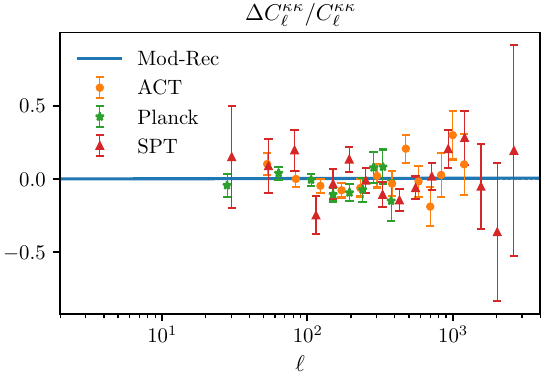}
     \caption{
     Relative difference in the CMB lensing convergence spectrum $C_\ell^{\kappa \kappa}$ for a modified recombination model with an $r_d$ smaller by 2\%, relative to the Planck best-fit $\Lambda$CDM model. Orange points, green stars, and red triangles with error bars show the $C_\ell^{\kappa \kappa}$ bandpowers from \textit{ACT}, \textit{Planck}, and \textit{SPT-3G}, respectively. The figure illustrates that recombination changes capable of raising $H_0$ to $\sim 72$ km/s/Mpc have only a minor effect on the CMB lensing spectrum.}
    \label{fig:rec_indep}
\end{figure}

\begin{figure*}[htbp!]
    \centering
    \includegraphics[width=0.90\textwidth]{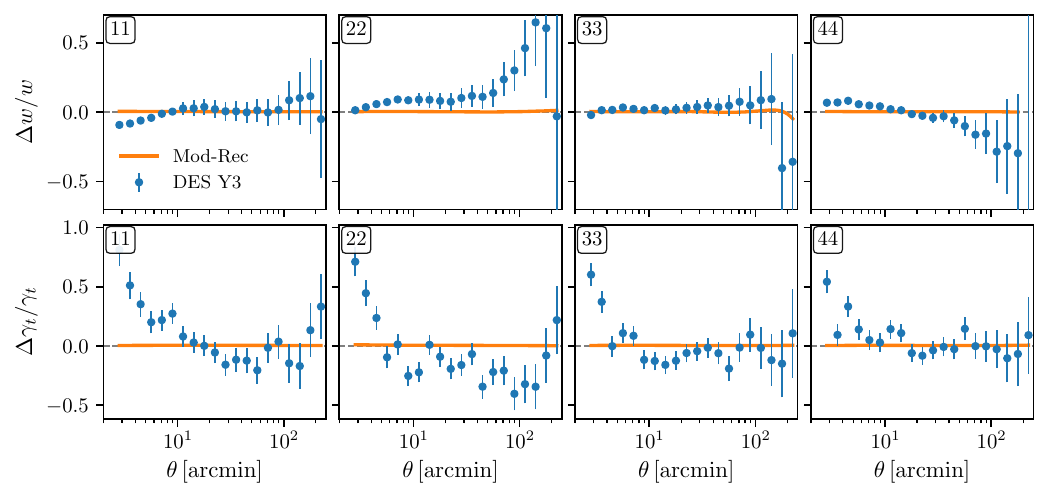}
     \caption{
     Relative differences in the galaxy clustering correlation functions $w_{ij}(\theta)$ and $\gamma_{t,ij}(\theta)$, in four redshift bins ($i=j=1,\dots,4$), with respect to the prediction of the Planck+DES~Y3 best-fit $\Lambda$CDM model. Blue points show the DES~Y3 measurement residuals with uncertainties, and the solid yellow curve corresponds to the same modified-recombination model as in Fig.~\ref{fig:rec_indep}. The figure demonstrates that the DES~Y3 $3\times2$\,pt observables are essentially insensitive to modifications of the recombination history.}
    \label{fig:des_res}
\end{figure*}

The late-time matter distribution can also be probed through galaxy clustering and galaxy weak lensing, which depend on both the amplitude and scale-dependence of matter fluctuations. These observables provide complementary information to CMB lensing. The three two-point correlation functions ($3\times2$-pt) are

\begin{align}
    w_{ij}(\theta) &= \langle \delta_g^i(\hat{n}) \, \delta_g^j(\hat{n}+\theta) \rangle, \label{eq:gg}\\
    \gamma_{t,ij}(\theta) &= \langle \delta_g^i(\hat{n}) \, \gamma_t^j(\hat{n}+\theta) \rangle, \label{eq:gl}\\
    \xi_{\pm,ij}(\theta) &= \langle \gamma^i(\hat{n}) \, \gamma^j(\hat{n}+\theta) \rangle, \label{eq:ss}
\end{align}

where $\delta_g^i$ is the galaxy overdensity in redshift bin $i$, $\gamma_t^j$ is the tangential shear in bin $j$, and $\gamma^i$ is the shear field. In this work, we include all three correlation functions.

To assess the sensitivity of CMB lensing and galaxy $3\times2$-pt data to the sound horizon, we compared observables computed with the standard ionization history \( x_e(z) \) in the Planck best-fit \(\Lambda\)CDM cosmology to those obtained with the same cosmological parameters but a modified \( x_e(z) \). Specifically, we used the four-parameter phenomenological model of~\citep{Mirpoorian:2024fka}, which yields an \( r_d \) smaller by about 2\%, corresponding to $H_0 \sim 72$ km/s/Mpc. Figure~\ref{fig:rec_indep} shows the resulting relative difference in \( C_\ell^{\kappa\kappa} \), alongside lensing bandpowers from Planck, ACT-DR6, and SPT-3G. The change remains below 1\% and well within current measurement uncertainties. Similarly, Fig.~\ref{fig:des_res} shows the residuals in $w_{ij}(\theta)$ and $\gamma_{t,ij}(\theta)$, compared to DES~Y3 measurements~\citep{DES:2021bwg,DES:2021wwk}. For brevity, $\xi_{\pm,ij}(\theta)$ is not shown, as it also exhibits no significant difference. These comparisons demonstrate that CMB lensing and galaxy $3\times2$-pt observables are effectively insensitive to moderate reductions in $r_d$ induced by modified recombination physics.

\subsection{The effect of massive neutrinos}
\label{sec:neutrinos}

Massive neutrinos affect both the expansion history and the growth of cosmic structures. Neutrinos with total mass $\Sigma m_{\nu} \lesssim 1$~eV behave as relativistic species before photon–baryon decoupling, contributing to the radiation energy density at $z \gtrsim 1000$ and delaying matter–radiation equality. This alters the horizon scale at equality, $ k_{\rm eq} $, imprinted in the matter power spectrum. On smaller scales, neutrino free-streaming suppresses the growth of density fluctuations below the free-streaming length, producing a scale-dependent suppression of the matter power spectrum. Both effects influence the CMB lensing convergence spectrum, modifying its amplitude and shape, as shown in Fig.~2 of~\citep{Lynch:2025ine}. Combining CMB lensing with other cosmological probes that constrain the present-day matter density fraction therefore provides sensitivity to the neutrino mass (see, e.g.,~\citep{Loverde:2024nfi,Lynch:2025ine,Craig:2024tky}).

The $3\times2$-pt galaxy weak lensing and clustering data, such as DES measurements, constrain the clustering amplitude and expansion history at low redshift, strengthening joint inference and significantly reducing the allowed parameter space for neutrino masses. Massive neutrinos suppress small-scale power and lower the amplitude of clustering and shear correlations, while the altered growth history modifies their redshift evolution. Because galaxy weak lensing and clustering respond differently to changes in the growth of structure and in $\Omega_m$, their combination provides complementary sensitivity to $\Sigma m_\nu$ through both scale-dependent suppression and changes in $\Omega_m$.

\subsection{Parameter degeneracies}
\label{degeneracies}

In \(\Lambda\)CDM, BAO constrain \( r_d H_0 \) and \( \Omega_m \), while the CMB lensing spectrum amplitude constrains approximately (see Eq.~(2.10) of \citep{Craig:2024tky})
\begin{equation}
C_\ell^{\kappa\kappa} \propto (\Omega_m h^2)^2 A_s 
\left[ 1 - 0.003 \frac{\Sigma m_\nu / (58~\mathrm{meV})}{\Omega_m h^2} \right].
\end{equation}
This scaling shows that, even if the neutrino fraction is known, external information on $A_s$ is needed to determine $\Omega_m h^2$. With $\Sigma m_\nu$ free, the CMB lensing amplitude alone cannot disentangle $\Omega_m h^2$, $A_s$, and $f_\nu$. An additional constraint on $\Sigma m_\nu$ comes from the scale-dependent suppression of $C_\ell^{\kappa\kappa}$, illustrated in Fig.~2 of~\citep{Lynch:2025ine}. The detectability of this suppression depends on the noise in $C_\ell^{\kappa\kappa}$. Including galaxy lensing and clustering extends the mode coverage and enhances the ability to disentangle $\Omega_m h^2$, $A_s$, and $\Sigma m_\nu$. In addition, \( A_s \) is well constrained by large-scale CMB anisotropies, which are independent of $r_d$, justifying the adoption of an informative prior on \( A_s \) from CMB data (see Sec.~\ref{sec:priors}). 

Allowing for massive neutrinos, which contribute as radiation at equality but as part of \( \Omega_m \) today, introduces a positive correlation between \( \Sigma m_\nu \) and \( H_0 \) when using CMB lensing and BAO. A larger \( \Sigma m_\nu \) reduces the effective matter density at equality, requiring a larger \( \Omega_m h^2 \) to maintain the same \( k_{\rm eq} \). With \( \Omega_m \) constrained by BAO, this in turn requires a larger \( H_0 \). These parameter relationships are evident in Fig.~\ref{fig:triangle_current_all}, which shows the joint posteriors for \( H_0 \), \( \Sigma m_\nu \), and \( \Omega_m h^2 \) (see also Fig.~2 of \citep{Lynch:2025ine}).

\section{Analysis details}

\subsection{Datasets}
\label{sec:lik}

\subsubsection{DESI DR2 BAO}

We use the DESI DR2 BAO {\tt Cobaya} likelihood, based on measurements provided in Table~III of \citep{DESI:2025zgx}. The dataset includes the $D_V/r_d$ measurement from the BGS sample at $0.1<z<0.4$, $D_M/r_d$ and $D_H/r_d$ for two LRG bins at $0.4<z<0.6$ and $0.6<z<0.8$, a combined LRG+ELG tracer at $0.8<z<1.1$, ELG measurements at $1.1<z<1.6$, QSO clustering BAO at $0.8<z<2.1$, and the Ly$\alpha$ BAO constraint at $1.8<z<4.2$. We refer to this dataset as DESI2.

\subsubsection{The CMB acoustic scale $\theta_{\star}$}
\label{thetaCMB}

The CMB acoustic scale angle, \( \theta_{\star} \), also known as \( \theta_{\mathrm{CMB}} \), is the angular size of the sound horizon at the epoch of recombination. We use \( \theta_{\star} \) as a transverse BAO measurement, \( \beta^{\star}_\perp = 94.286 \pm 0.217 \), at \( z_{\star} = 1090 \)~\citep{Pogosian:2024ykm}, derived from \( \theta_{\star} \) measured by Planck under the assumption that the \(\Lambda\)CDM relation between \( r_d \) and the sound horizon at the redshift corresponding to the peak of the CMB visibility function, \( r_* \), holds (with \( r_{\star} \approx r_d / 1.02 \)). The authors verified that this relation remains valid in known extensions of \(\Lambda\)CDM that modify \( r_d \).

Adding $\theta_{\star}$ to our dataset helps tighten the BAO constraints on $r_dH_0$ and $\Omega_m$.

\subsubsection{CMB Lensing: \textit{Planck}, ACT, and SPT-3G}

We use the combination of CMB lensing bandpowers from \textit{Planck}~\citep{Planck:2020olo}, the Atacama Cosmology Telescope (ACT DR6)~\citep{ACT:2023dou,ACT:2023ubw, ACT:2023kun}, and the South Pole Telescope (SPT-3G)~\citep{SPT-3G:2024atg} provided in \citep{SPT-3G:2025zuh}. This combination, subsequently referred to as APS-L, provides the most constraining CMB lensing power spectrum measurement to date with a combined signal-to-noise ratio of 61, offering strong constraints on the amplitude of matter fluctuations $S_8$~\citep{SPT-3G:2025zuh}.

\subsubsection{Dark Energy Survey $3\times2$-pt Correlation Functions}

We include the full $3\times2$-pt correlation functions from the Dark Energy Survey Year 3 (DES Y3) encompassing cosmic shear, galaxy clustering, and galaxy-galaxy lensing as detailed in~\citep{DES:2021wwk}. This joint analysis over $\sim5000\ \mathrm{deg}^2$ provides competitive constraints on cosmological parameters, yielding $S_8 = 0.776^{+0.017}_{-0.017}$ and $\Omega_m = 0.339^{+0.032}_{-0.031}$, and serves as a powerful low-redshift complement to CMB-based probes. By directly tracing the growth of structure, DES Y3 measurements are particularly sensitive to the suppression effects induced by massive neutrinos, thereby providing valuable information on the sum of neutrino masses. 

We use the \texttt{Cobaya}~\citep{Torrado:2020dgo} adaption of the DES-Y3 3x2-pt likelihood provided in \citep{Wang:2024uvw}. In particular, we use the configuration with the Tidal Alignment and Tidal Torquing (TATT) intrinsic alignment model (with shear-ratio included), validated to reproduce the official DES Y3 results~\citep{DES:2021wwk}.

\subsubsection{Pantheon+ Supernovae}

We use the Pantheon+ (PP) dataset of 1550 spectroscopically-confirmed Type Ia SN spanning \( 0.001 < z < 2.26 \) to probe the late-time expansion history \citep{Scolnic:2021amr,Brout:2022vxf}. This updated compilation improves upon the original Pantheon release through enhanced photometric calibration, extended low-redshift coverage, and refined treatment of systematics.

\subsection{Model Parameters and Priors} 
\label{sec:priors}

We use the Monte-Carlo Markov Chains (MCMC) package \texttt{Cobaya}~\citep{Torrado:2020dgo} with the Boltzmann code \texttt{CAMB}~\citep{Lewis:1999bs} modified to treat \( r_d \) as a primary (rather than derived) parameter when computing the BAO observables. Specifically, in \texttt{CAMB}, we introduce a new primary parameter \texttt{rd} and modify the BAO likelihood file of Cobaya (\texttt{bao.py}) to use it instead of the default derived sound horizon parameter \texttt{rdrag}. Computation of observables other than BAO is not modified.

The sampled parameters are \( r_d \), \( H_0 \), \( \Omega_c h^2 \), \( \Omega_b h^2 \), the spectral index \( n_s \), \( A_s \), and \( \Sigma m_\nu \). We also consider the case where \( \Sigma m_\nu \) is fixed at the minimal value of 0.06~eV. The parameter priors adopted in this work are summarized in Table~\ref{tab:priors}. Below, we briefly explain the reasoning behind our choices of priors. 

\begin{table}[htbp!]
    \begin{ruledtabular}
    \centering
    \begin{tabular}{|c|cc|}
 & Current Data & Forecast\\
 & Prior & Fiducial value, Prior\\
         \colrule
         \rule{0pt}{3ex}
         $\ln(10^{10} A_s)$ & [2, 4] or $\mathcal{N}(3.04, 0.02)$ & $3.045$, $\mathcal{N}(3.04, 0.02)$ \\
         $H_0$ {[km/s/Mpc]} & {[40, 100]}  & 67.32,  {[40, 100]} \\
         $n_s$ & $\mathcal{N}(0.96, 0.02)$ &  0.966, $\mathcal{N}(0.96, 0.02)$ \\
         100 $\Omega_b h^2$ &  $\mathcal{N}(2.23, 0.05)$ & 2.24, $\mathcal{N}(2.23, 0.05)$ \\
         $\Omega_c h^2$ & [0.001, 0.99]  & 0.120, [0.001, 0.99] \\
         $\Sigma m_{\nu}$ [eV] & [0, 3.0]  & 0.06, [0, 3.0] \\
         $r_d$ [Mpc] & [100, 200] & 147.1, [100, 200]\\
         $\tau$ (fixed) & 0.059  & 0.059
    \end{tabular}
    \end{ruledtabular}
    \caption{Priors adopted for the parameters in the analysis of current data, and the fiducial values and priors used in the forecast. Uniform priors are indicated by square brackets, while Gaussian priors with mean $\mu$ and standard deviation $\sigma$ are denoted as $\mathcal{N}(\mu, \sigma)$.}
    \label{tab:priors}
\end{table}

For the $H_0$ and the cold dark matter density parameter, $\Omega_c h^2$, we use flat uninformative priors with ranges given in Table~\ref{tab:priors}.

When analysing the current data, we present results both with an uninformative flat prior on $A_s$, $\ln(10^{10} A_s) \in [2, 4]$ (same as in \citep{SPT-3G:2025zuh}), as well a Gaussian prior, $\mathcal{N}(3.04, 0.02)$, centered on the Planck-$\Lambda$CDM preferred value~\citep{Carron:2022eyg} with a standard deviation encompassing the $A_s$ posteriors obtained in a broad range of modified recombination models~\citep{Mirpoorian:2024fka}. In particular, the modified recombination model from~\citep{Mirpoorian:2024fka} that substantially relieves the Hubble tension, with $H_0 = 72.57 \pm 0.47$ km/s/Mpc, has $\ln(10^{10} A_s) = 3.042 \pm 0.014$. The $A_s$ posterior in Early Dark Energy models, $\ln(10^{10} A_s) = 3.067 \pm 0.017$,~\citep{Chaussidon:2025npr} is also within the 1-$\sigma$ range of our prior. We note that a similar Gaussian prior of $\ln(10^{10} A_s) = 3.0372 \pm 0.0194$ was used in a recent $r_d$-agnostic study~\citep{Chaussidon:2025npr} that appeared shortly after the submission of our paper.

Having a stronger prior on $A_s$ helps break the degeneracy between $A_s$ and $\Omega_mh^2$ in CMB lensing, significantly tightening the posterior of $H_0$.
  
For the scalar spectral index $n_s$, we adopt a Gaussian prior $\mathcal{N}(0.96, 0.02)$, same as the prior used in \citep{SPT-3G:2025zuh}, that encompasses a broad range of posteriors in models with modified recombination histories. We found that adopting a more restrictive prior on $n_s$ did not appreciably alter the constraints on the parameters of interest.

Varying the optical depth to reionization, $\tau$, has no impact on our posterior distributions. We therefore fix this parameter to the Planck + DESI DR2  $\Lambda$CDM best-fit value, $\tau = 0.059$ \citep{Mirpoorian:2025rfp}. This choice is also consistent with the value obtained from the {\tt SRoll2} low-$\ell$ polarization analysis~\citep{Pagano:2019tci}. Given the key role of $\tau$ for the parameter constraints derived from the combination of CMB temperature, polarization and lensing spectra, with implications for the apparent tension between the Planck CMB and DESI BAO~\citep{Sailer:2025lxj}, having a $\tau$-insensitive way of measuring a subset of cosmological parameters offers a valuable cross-check.

We adopt a Gaussian prior on the physical baryon density, $\Omega_b h^2 \sim \mathcal{N}(0.0223, 0.0005)$, obtained from observations of the primordial abundances of light elements, particularly deuterium, and is consistent with the standard BBN scenario combined with updated nuclear reaction rate data \citep{Cooke:2017cwo}.

We adopt a wide and minimally informative flat prior on the net neutrino mass, $\Sigma m_{\nu}$, spanning $[0, 3]$~eV, allowing our constraints to remain largely driven by the data~\citep{Escudero:2024uea}. We also perform separate analyses with $\Sigma m_{\nu}$ fixed at the minimum value of $0.06$~eV determined by neutrino oscillation experiments~\citep{KamLAND:2002uet,NOvA:2004blv,SNO:2002tuh,Esteban:2020cvm,deSalas:2020pgw}. We note that prior choice for the neutrino mass sum has been shown to significantly affect Bayesian evidence and posterior inferences, especially in the case of significant parameter degeneracies. Several works have emphasized the impact of adopting linear versus logarithmic priors, particularly in distinguishing between neutrino mass orderings and assessing constraints on $\Sigma m_\nu$~\citep{Hergt:2021qlh, Gariazzo:2018pei, Long:2017dru, Gariazzo:2022ahe, Simpson:2017qvj}. In light of these findings, our baseline analysis adopts a uniform prior on $\Sigma m_\nu$, consistent with common practice in cosmological analyses. However, we also demonstrate in Fig.~\ref{fig:log_prior_mnu} the impact of using a logarithmic prior. In addition, a number of groups have explored analyses where $\Sigma m_\nu$ is allowed to take negative values, which can be informative for isolating parameter degeneracies and quantifying the statistical power of cosmological data~\citep{Craig:2024tky,Elbers:2024sha, Lynch:2025ine}. Investigating negative-mass parameterizations lies beyond the scope of this project.

\subsection{Forecasts}
\label{ss:mock}

In what follows, we describe the methodology used in our forecasts of the constraints expected from future CMB lensing, galaxy lensing and clustering, BAO and SN datasets. We adopt the Planck best-fit $\Lambda$CDM model as our fiducial cosmology with parameters provided in Table~\ref{tab:priors}.

\subsubsection{CMB Lensing}

For the forecast, we construct a lensing likelihood function $\mathcal L$ assuming Gaussian errors and covariance for the binned CMB lensing convergence spectrum, $\hat{C_{L_b}^{\kappa\kappa}}$ \citep{ACT:2023dou, ACT:2023kun}:
\begin{equation}
    - 2 \ln \mathcal{L} = \Sigma_{bb'} \left[ \hat{C}_{L_b}^{\kappa\kappa} - C_{L_b}^{\kappa\kappa}(\theta) \right] \mathbb{C}_{bb'}^{-1} \left[ \hat{C}_{L_b'}^{\kappa\kappa} - C_{L_b'}^{\kappa\kappa}(\theta) \right],
    \label{eq:likelihood}
\end{equation}
where $b$ denotes a bin centered at multiple $L_b$, $C_{L_b}^{\kappa\kappa}(\theta)$ is the theory spectrum calculated using a set of cosmological parameters $\theta$, $\hat{C}_{L_b}^{\kappa\kappa}$ is the mock ``measured'' spectrum and $\mathbb{C}_{bb'}$ is the covariance matrix generated as described below.

We adopt the extended range bin configuration used in the ACT DR6 analysis, with the $N_{\rm bins} = 18$ non-overlapping bins centered at $L =$ [14, 30, 53, 84, 123, 172, 232, 302, 382, 476, 582, 700, 832, 1001, 1200, 1400, 1600, 1874]. To evaluate the impact of using a different binning scheme on the forecasted parameter constraints, we also generated mock CMB lensing datasets $N_{\rm bins}$ ranging from $13$ to $100$, with the bin widths forming an arithmetic progression resulting in progressively wider bins at higher multipoles. We found that the uncertainties in the cosmological parameters converged for the 18-bin case we have adopted.

To generate a single realization of the mock data, we use the additive Gaussian noise approximation by drawing a random sample from a Gaussian distribution centered at the fiducial cosmology $C^{\kappa\kappa}_L$ and standard deviation given by
\begin{equation}
    \Delta C^{\kappa\kappa}_L = \sqrt{\frac{2}{f_{\rm sky} (2L + 1)}} \left( C^{\kappa\kappa}_L + N^{\kappa\kappa}_L\right),
    \label{eq:Delta_cl}
\end{equation}
where $N^{\kappa\kappa}_L$ is the noise power spectrum, and $f_{\rm sky}$ is the fraction of sky coverage \citep{Knox:1995dq}. $N^{\kappa\kappa}_L$ receives contributions from all sources of small-scale CMB fluctuation power, including primary CMB anisotropies, detector noise, atmospheric noise, and any other instrumental or astrophysical contributions that enter the reconstruction. Following the methodology described in \citep{ACT:2023dou}, we obtain the band-power covariance matrix from $N_{\rm mock} = 796$ mock simulations while assuming SO-like sky coverage $f_{\rm sky} = 0.5$. To account for the biased estimate of the covariance matrix from a finite number of realizations, we rescale the estimated inverse covariance matrix by the Hartlap factor \citep{Hartlap:2006kj}:
\begin{equation}
    f_{\rm H} = \frac{N_{\rm mock} - N_{\rm bins} - 2}{N_{\rm mock} - 1},
    \label{eq:hartlap}
\end{equation}
where $N_{\rm bins}$ is the number of bandpowers.

As the noise power spectrum $N^{\kappa\kappa}_L$, we use the Simons Observatory minimum variance (MV) noise curve {\tt v3.1.1} with the goal sensitivity for the large aperture telescope (LAT) \citep{SimonsObservatory:2018koc}. The dominant contribution arises from the disconnected Gaussian term, $N^{(0)}$, which originates from random correlations in the primary CMB anisotropies \citep{Madhavacheril:2020ido}. Additional noise comes from instrumental and atmospheric fluctuations, which introduce anisotropic variance in the maps. Foreground contamination, including the thermal and kinetic Sunyaev–Zel’dovich effects, cosmic infrared background, and unresolved point sources, adds non-Gaussian structure that biases lensing estimators. Furthermore, masking of bright sources and survey geometry effects can induce additional biases if not properly accounted for \citep{Atkins:2023yzu}.

\subsubsection{$3\times2$-pt galaxy weak lensing and clustering}

To model a future galaxy weak lensing and clustering dataset ($w_{ij}$, $\gamma_{t,ij}$, and $\xi_{\pm,ij}$), we generate mock DES-Year-1-like 3$\times$2-pt data vectors based on our fiducial $\Lambda$CDM cosmology. We then add Gaussian noise drawn from the DES Y1 covariance matrix, rescaled by a factor of 25 to approximate the precision expected from future surveys. The factor of 25 is a rough estimate based on the expected increase in the galaxy number density from approximately 5.5 gal/arcmin$^2$ to 26–30 gal/arcmin$^2$ for LSST, and the increase in the sky coverage from 3.2\% for DES Y1 to upto 44\% for LSST.

Our mock likelihood is based on the publicly available {\tt Cobaya} implementation of the DES Y1~\citep{DES:2017tss, DES:2018ufa} likelihood, with the rescaled covariance and the mock data.

For the MCMC runs with the mock SO-L CMB lensing data, we vary and marginalize over the DES-Y1 likelihood nuisance parameters. For runs combined with the CV-L CMB lensing mock, we fixed the nuisance parameters at their fiducial values. The former setup is meant to represent a combination that can be expected to be available in the relatively near future, while the latter is more representative of the ultimate limit. Throughout the paper, we denote the 3$\times$2-pt mock analysis with fixed nuisance parameters with an asterisk, {\it i.e.} $3\times2$-pt$^*$, in both tables and figures.

\subsubsection{BAO \& SN forecast methodology}

To model a future BAO dataset, we generate mock data for the three types of BAO observables $D_M/r_d, D_H/r_d$, and $D_V/r_d$ based on the fiducial model and the DESI DR2 data covariance, reduced by a factor of 2. This is meant to approximate the BAO data from the completed full 14,000 deg$^2$ DESI program~\citep{DESI:2023dwi}, which will cover approximately twice the volume included in the DR2 release. We obtain a covariance matrix from $N_{\rm mock} = 796$ simulations, with the mock BAO estimates given as 13 data points at 7 effective redshifts: $z_{\rm eff} =$ [0.295, 0.510, 0.706, 0.934, 1.321, 1.484, 2.330], that match the redshifts in the native {\tt Cobaya} DESI DR2 BAO likelihood and making it easy to modify it for use with our mock data.

To model a future SN~Ia dataset, we adopt the Pantheon+ sample~\citep{Brout:2022vxf} as our template, computing the theoretical distance modulus, $\mu_{\rm th}(z)$, at each redshift using {\tt CAMB}~\citep{Lewis:1999bs} for the Planck $\Lambda$CDM best-fit parameters. Mock data are then generated by adding Gaussian noise with standard deviations given by the Pantheon+ uncertainties reduced by a factor of $2.5$, corresponding to an ``ultimate'' SN sample with roughly six times more objects than the current dataset, as expected from Vera Rubin LSST and Euclid, both of which forecast around $10^4$ high-quality SN Ia~\citep{LSSTDarkEnergyScience:2018jkl,Astier:2014swa}. The mock SN sample is then binned into $N_{\rm bins} = 40$ redshift bins using an equal-occupancy (quantile) scheme, which ensures that each bin contains a comparable number of SN and thus similar statistical precision on $\mu(z_b)$.  
For validation, we also tested an alternative ``Union3'' binning scheme with fixed bin centers matching the Union3~\citep{Rubin:2023jdq} compilation.

We produce $N_{\rm mock} = 796$ independent realizations of the binned data vector $\hat{\mu}(z_b)$ and estimate the covariance matrix $\mathbb{C}_{bb'}$ from these realizations. To correct for the bias in the inverse covariance from a finite number of realizations, we rescale it by the Hartlap factor.

The resulting mock SN likelihood has the form
\begin{equation}
    - 2 \ln \mathcal{L} = \sum_{bb'} 
    \left[ \hat{\mu}(z_b) - \mu_{\rm th}(z_b; \theta) \right] 
    \mathbb{C}^{-1}_{bb'}
    \left[ \hat{\mu}(z_{b'}) - \mu_{\rm th}(z_{b'}; \theta) \right],
\end{equation}
where $\theta$ denotes the set of cosmological parameters. 

This approach yields a mock SN likelihood that retains the statistical properties and redshift coverage of Pantheon+, but with the improved constraining power expected from Vera Rubin LSST and Euclid.

\section{Results}
\label{sec:results}

\begin{table*}[htbp!]
\centering
\begin{tabular}{c|cccccc}
\hline
\hline
{\bf Current Data} (fixed $\Sigma m_\nu$)
& $H_0$ [km/s/Mpc]
& $100\,\Omega_m h^2$
& $\Omega_m$
& $r_d$ [Mpc]
& $\ln(10^{10} A_s)$
& $S_8$ \\
\hline
\footnote{In the analysis using the full Planck data, $r_d$ is a derived parameter computed using the standard recombination routine in {\tt CAMB}. In the analyses of all other data combinations in this table, $r_d$ is a free parameter.}
DESI2+Planck+APS-L+DESY3+PP
& $68.0 \pm 0.25$
& $14.11\pm0.05$
& $0.305 \pm 0.003$
& $147.7 \pm 0.17$
& $3.060 \pm 0.014$
& $0.819 \pm 0.007$ \\
DESI2+$\theta_{\star}$+APS-L (Base)
& $70.5 \pm 2.0$
& $14.75\pm0.81$
& $0.297 \pm 0.005$
& $144.2 \pm 4.0$
& $3.039 \pm 0.073$
& $0.831 \pm 0.011$ \\
Base+$A_s$
& $70.42 \pm 0.99$
& $14.71 \pm 0.30$
& $0.297 \pm 0.005$
& $144.3 \pm 1.6$
& $3.040 \pm 0.019$
& $0.830 \pm 0.011$ \\
Base+DESY3
& $70.3 \pm 1.7$
& $14.62 \pm 0.64$
& $0.295 \pm 0.005$
& $144.7 \pm 3.2$
& $3.045 \pm 0.059$
& $0.826 \pm 0.009$ \\
Base+DESY3+$A_s$
& $70.40 \pm 0.98$
& $14.65 \pm 0.29$
& $0.296 \pm 0.005$
& $144.5 \pm 1.6$
& $3.041 \pm 0.019$
& $0.826 \pm 0.009$ \\
Base+DESY3+PP
& $70.0 \pm 1.7$
& $14.60^{+0.60}_{-0.68}$
& $0.298 \pm 0.005$
& $145.0 \pm 3.2$
& $3.044 \pm 0.060$
& $0.827 \pm 0.009$ \\
Base+DESY3+PP+$A_s$
& $70.03 \pm 0.97$
& $14.62 \pm 0.29$
& $0.298 \pm 0.005$
& $144.8 \pm 1.6$
& $3.041 \pm 0.019$
& $0.827 \pm 0.009$ \\
\hline
\hline
{\bf Forecast} (fixed $\Sigma m_\nu$)
& & & & & & \\
\hline
BAO+$\theta_{\star}$+SO-L+SN+$A_s$
& $67.40 \pm 0.71$
& $14.33 \pm 0.23$
& $0.315 \pm 0.004$
& $146.8 \pm 1.3$
& $3.037 \pm 0.019$
& $0.830 \pm 0.004$ \\
BAO+$\theta_{\star}$+SO-L+$3\times2$-pt+SN
& $67.30^{+1.1}_{-1.2}$
& $14.33^{+0.43}_{-0.51}$
& $0.316 \pm 0.003$
& $146.9 \pm 2.5$
& $3.037 \pm 0.045$
& $0.831 \pm 0.003$ \\
BAO+$\theta_{\star}$+SO-L+$3\times2$-pt+SN+$A_s$
& $67.24 \pm 0.67$
& $14.30 \pm 0.22$
& $0.316 \pm 0.003$
& $147.0 \pm 1.2$
& $3.039 \pm 0.019$
& $0.831 \pm 0.003$ \\
BAO+$\theta_{\star}$+CV-L+SN+$A_s$
& $67.30 \pm 0.70$
& $14.37 \pm 0.22$
& $0.317 \pm 0.004$
& $147.2 \pm 1.3$
& $3.040 \pm 0.019$
& $0.835 \pm 0.002$ \\
BAO+$\theta_{\star}$+CV-L+$3\times2$-pt$^*$+SN
& $67.10 \pm 1.0$
& $14.27 \pm 0.41$
& $0.317 \pm 0.002$
& $147.8 \pm 2.2$
& $3.051 \pm 0.038$
& $0.835 \pm 0.002$ \\
BAO+$\theta_{\star}$+CV-L+$3\times2$-pt$^*$+SN+$A_s$
& $67.30 \pm 0.57$
& $14.35 \pm 0.21$
& $0.317 \pm 0.002$
& $147.3 \pm 1.2$
& $3.042 \pm 0.018$
& $0.834 \pm 0.002$ \\
\hline
\end{tabular}
\caption{Constraints on cosmological parameters from current data, assuming fixed $\Sigma m_\nu = 0.06$~eV. The “Base” dataset denotes DESI2+$\theta_\star$+APS-L. Uncertainties are quoted at 68\% CL.}
\label{tab:fixmnu}
\end{table*}

\begin{table*}[!htbp]
\centering
\begin{tabular}{c|cccccc}
\hline
\hline
{\bf Current Data} (varying $\Sigma m_\nu$)
& $H_0$ [km/s/Mpc]
& $100\,\Omega_m h^2$
& $\Omega_m$
& $\Sigma m_\nu$ [eV]
& $\ln(10^{10} A_s)$
& $S_8$ \\
\hline
\footnote{In the analysis using the full Planck data, $r_d$ is a derived parameter computed using the standard recombination routine in {\tt CAMB}. In the analyses of all other data combinations in this table, $r_d$ is a free parameter.}
DESI2+Planck+APS-L+DESY3+PP
& $68.21 \pm 0.27$
& $14.09 \pm 0.06$
& $0.303 \pm 0.003$
& $<0.022$ (0.050)
& $3.051 \pm 0.013$
& $0.823 \pm 0.007$ \\
Base+DESY3
& $75.7^{+3.1}_{-4.2}$
& $17.0^{+1.3}_{-2.0}$
& $0.297 \pm 0.005$
& $0.54^{+0.23}_{-0.38}$ (1.10)
& $3.007 ^{+ 0.073}_{-0.065}$
& $0.821 \pm 0.009$ \\
Base+DESY3+$A_s$
& $74.2 \pm 2.2$
& $16.32 \pm 0.96$
& $0.297 \pm 0.005$
& $0.45^{+0.20}_{-0.27}$ (0.862)
& $3.038 \pm 0.019$
& $0.821 \pm 0.009$ \\
Base+DESY3+PP
& $75.3^{+3.3}_{-4.0}$
& $17.0^{+1.4}_{-1.9}$
& $0.299 \pm 0.005$
& $0.55^{+0.23}_{-0.37}$ (1.11)
& $3.006 \pm 0.069$
& $0.822 \pm 0.009$ \\
Base+DESY3+PP+$A_s$
& $73.9 \pm 2.2$
& $16.36 \pm 0.92 $
& $0.299 \pm 0.005$
& $0.46^{+0.21}_{-0.25}$($0.46^{+0.40}_{-0.45}$)
& $3.038 \pm 0.019$
& $0.824 \pm 0.010$ \\
Base+DESY3 ($\log \Sigma m_\nu$)
& $71.9^{+1.8}_{-3.5}$
& $15.31^{+0.66}_{-1.5}$
& $0.296 \pm 0.005$
& $0.196^{+0.079}_{-0.22}$($0.20^{+0.50}_{-0.23}$)
& $3.035 \pm 0.061$
& $0.825 \pm 0.010$\\
Base+DESY3+PP+$A_s$ ($\log \Sigma m_\nu$)
& $71.5^{+1.4}_{-2.9}$
& $15.26^{+0.53}_{-1.2}$
& $0.299 \pm 0.005$
& $0.205^{+0.097}_{-0.23}$($0.20^{+0.46}_{-0.24}$)
& $3.039 \pm 0.019$
& $0.826 \pm 0.009$ \\
\hline
\hline
{\bf Forecast} (varying $\Sigma m_\nu$) &&&&&& \\
\hline
BAO+$\theta_{\star}$+SO-L+SN+$A_s$
& $69.6^{+1.6}_{-2.3}$
& $15.32^{+0.64}_{-1.0}$
& $0.316 \pm 0.004$
& $<0.357$ (0.632)
& $3.037 \pm 0.019$
& $0.826 \pm 0.005$ \\
BAO+$\theta_{\star}$+SO-L+$3\times2$-pt+SN
& $67.8^{+1.2}_{-1.5}$
& $14.56^{+0.48}_{-0.64}$
& $0.317 \pm 0.003$
& $<0.135$ (0.273)
& $3.039 \pm 0.045$
& $0.830 \pm 0.004$ \\
BAO+$\theta_{\star}$+SO-L+$3\times2$-pt+SN+$A_s$
& $67.71^{+0.84}_{-1.1}$
& $14.52^{+0.31}_{-0.49}$
& $0.317 \pm 0.003$
& $<0.133$ (0.263)
& $3.040 \pm 0.018$
& $0.830 \pm 0.004$ \\
BAO+$\theta_{\star}$+CV-L+SN+$A_s$
& $68.40^{+0.98}_{-1.8}$
& $14.86^{+0.36}_{-0.76}$
& $0.317 \pm 0.004$
& $<0.188$ (0.403)
& $3.037 \pm 0.019$
& $0.833 \pm 0.004$ \\
BAO+$\theta_{\star}$+CV-L+$3\times2$-pt$^*$+SN
& $67.9^{+1.2}_{-1.4}$
& $14.62^{+0.47}_{-0.62}$
& $0.317 \pm 0.002$
& $0.122^{+0.050}_{-0.081}$ (0.247)
& $3.046 \pm 0.038$
& $0.833 \pm 0.003$ \\
BAO+$\theta_{\star}$+CV-L+$3\times2$-pt$^*$+SN+$A_s$
& $67.99^{+0.81}_{-0.97}$
& $14.67^{+0.33}_{-0.43}$
& $0.317 \pm 0.002$
& $0.123^{+0.051}_{-0.079}$($0.244$)
& $3.041 \pm 0.018$
& $0.833 \pm 0.003$ \\
\hline
\end{tabular}
\caption{ Constraints on cosmological parameters from current data, allowing $\Sigma m_\nu$ to vary. The “Base” dataset denotes DESI2+$\theta_\star$+APS-L. For $\Sigma m_\nu$, both 68\% CL intervals and 95\% CL upper bounds are shown. “($\log \Sigma m_\nu $)” denotes the results obtained using a logarithmic prior on $\Sigma m_\nu$ as shown in Figure~\ref{fig:log_prior_mnu}.}
\label{tab:varymnu}
\end{table*}

\begin{figure*}[htbp!]
    \centering
    \includegraphics[width=0.7\textwidth]{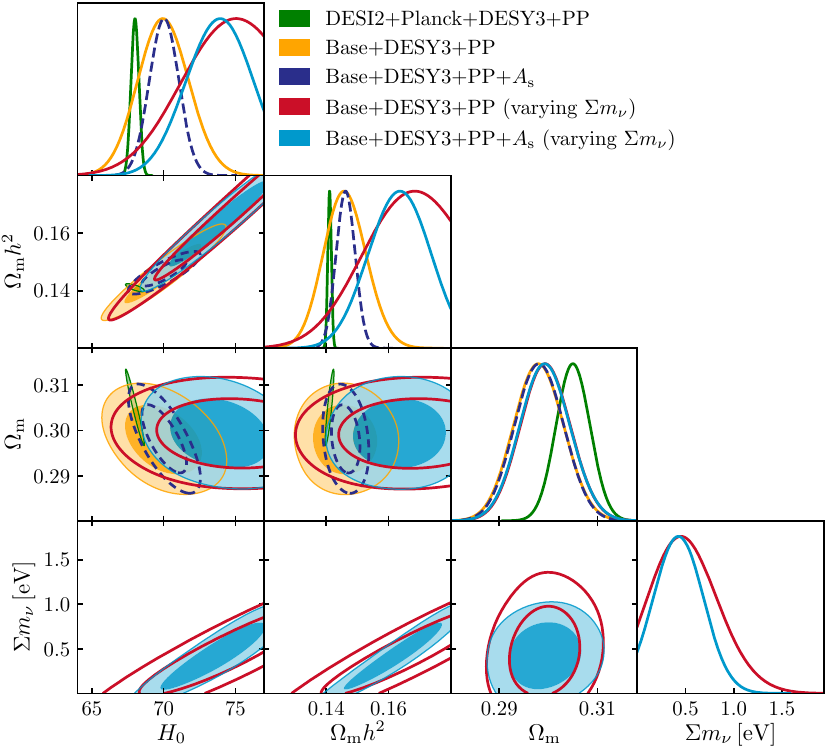}
    \caption{The 68\% and 95\% CL contours and the 1D marginalized posterior distributions for $H_0$, $\Omega_m h^2$, $\Omega_m$, and $\Sigma m_\nu$ from current data. The elongated 2D contours of $H_0$, $\Omega_m h^2$ and $\Sigma m_\nu$ highlight the positive correlations between these parameters discussed in the text. The DESI2+Planck+APS-L+DESY3+PP analysis uses the full Planck data, with $r_d$ as a derived parameter computed using the standard recombination routine in CAMB. In the analyses of all other data combinations in this figure, $r_d$ is a free parameter.}
    \label{fig:triangle_current_all}
\end{figure*}

In what follows, we present the $r_d$-agnostic parameter constraints inferred from the current data, as well as the forecasts. The results for the fixed and the varying $\Sigma m_\nu$ analyses are summarized in Tables~\ref{tab:fixmnu} and \ref{tab:varymnu}, respectively, and the key parameter posteriors are shown in Figures~\ref{fig:triangle_current_all} and \ref{fig:triangle_future_all}. The $r_d$-agnostic tests are compared to the standard $\Lambda$CDM fits to the combination of DESI2 BAO, PP SN and DES Year-3 with the full complement of CMB spectra, namely, the NPIPE {\tt CamSpec} high-$\ell$ likelihood \citep{Efstathiou:2019mdh,Rosenberg:2022sdy}, the PR3 low-$\ell$ TT and EE \citep{Planck:2018vyg,Planck:2019nip}, and CMB lensing measurements from Planck, ACT, and SPT-3G (APS-L)~\citep{SPT-3G:2025zuh} (DESI2+Planck+APS-L+DESY3+PP), where $r_d$ was a derived parameter computed using the standard recombination routine in {\tt CAMB}. Throughout this section, including in the tables and the figures, we refer to the DESI2+$\theta_\star$+APS-L combination as the ``Base'' dataset.

\begin{figure}[htbp!]
   \centering
   \includegraphics[width=0.48\textwidth]{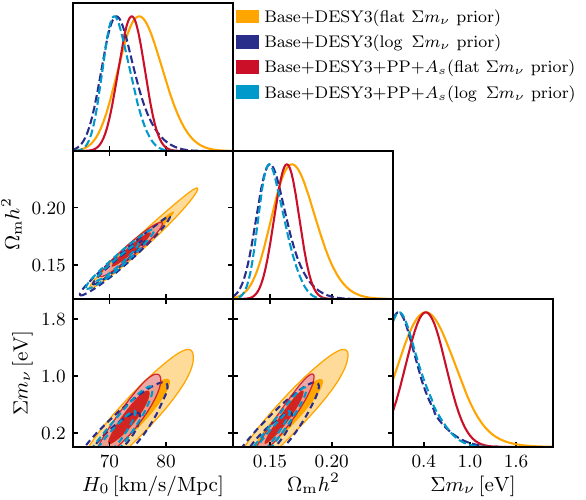}
   \caption{ The 68\% and 95\% CL contours, along with the 1D marginalized posterior distributions for $H_0$, $\Omega_m h^2$, and $\Sigma m_\nu$ from current data, comparing the results obtained with a uniform prior and a logarithmic prior on $\Sigma m_\nu$. This figure illustrates how using a logarithmic prior eliminates the apparent bias towards higher values of $H_0$ and $\Omega_mh^2$.}
   \label{fig:log_prior_mnu}
\end{figure}

\begin{figure*}[htbp!]
    \centering
    \includegraphics[width=0.7\textwidth]{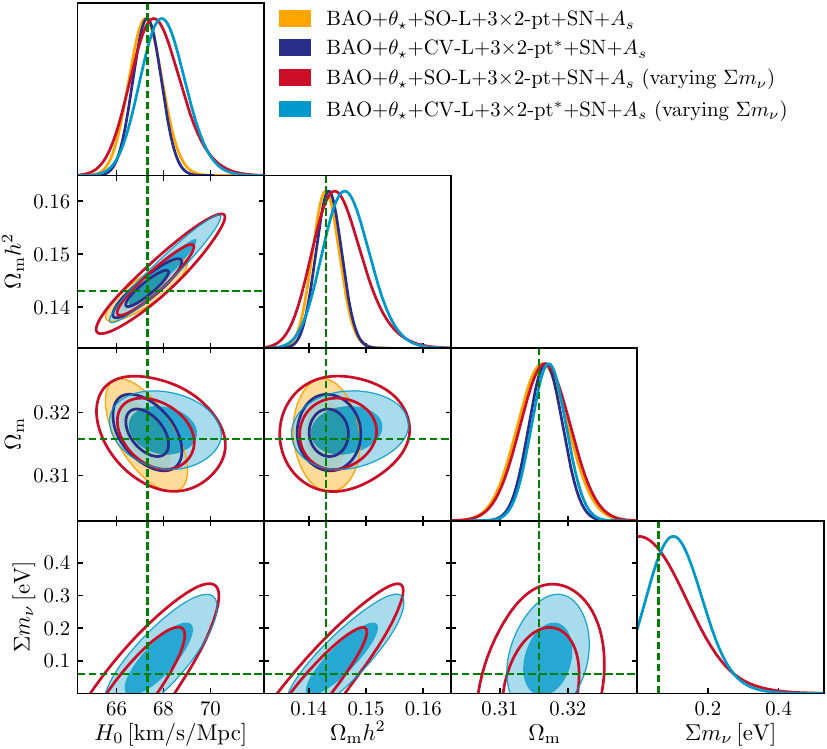}
    \caption{The forecasted 68\% and 95\% CL posterior distributions for $H_0$, $\Omega_m h^2$, $\Omega_m$, and $\Sigma m_\nu$. For reference, the green dashed lines represent the fiducial values used in the forecast.}
    \label{fig:triangle_future_all}
\end{figure*}

\subsection{Current data constraints with fixed $\Sigma m_\nu$}

With $r_d$ as a free parameter, the combination of DESI2, $\theta_{\star}$ and APS-L (Base) yields $H_0 = 70.5 \pm 2.0$~km/s/Mpc. Adding DESY3 improves the $H_0$ constraint by about 15\%, with $H_0 = 70.3 \pm 1.7$~km/s/Mpc.  Supplementing this with the CMB-based Gaussian prior on $A_s$ significantly reduces the uncertainty, resulting in $H_0 = 70.40 \pm 0.98$~km/s/Mpc, while also improving the constraints on $r_d$ and $\Omega_mh^2$. 

Adding the PP SN data leads to only minor shifts in the mean values and practically no reduction in parameter uncertainties. This lack of improvement despite SN offering an independent constraint on $\Omega_m$ can be explained by the fact that PP prefers a value of $\Omega_m$ \citep{Brout:2022vxf} that is $1.8\sigma$ higher than the one preferred by DESI2 \citep{DESI:2025zgx}. The combination of Base+DESY3+PP+$A_s$ yields $H_0 = 70.03 \pm 0.97$~km/s/Mpc that is $2\sigma$ above $H_0=68.0 \pm 0.25$~km/s/Mpc obtained from the conventional derived-$r_d$ fit to DESI2+Planck+APS-L+DESY3+PP, $2.4\sigma$ above the Planck-only value of $H_0=67.36 \pm 0.54$~km/s/Mpc and $2.1\sigma$ below the SH0ES Cepheid-based distance ladder measurement of $73.04 \pm 1.04$~km/s/Mpc.

The Base+DESY3+PP+$A_s$ value of $S_8=0.827 \pm 0.009$ is in good agreement with $S_8=0.825_{-0.013}^{+0.015}$ obtained in \citep{SPT-3G:2025zuh} from APS-L alone, and the DESI2+Planck+APS-L+DESY3+PP value of $S_8=0.819 \pm 0.007$.

In addition, we repeat the analysis of this section using the pre-DESI BAO data denoted as SDSS$^+$. This includes the eBOSS DR16 BAO measurements from LRGs, ELGs, QSOs, and the Lyman-$\alpha$ forest \citep{eBOSS:2020yzd}, the BOSS DR12 LRG BAO sample \citep{BOSS:2016wmc}, as well as low-redshift BAO from 6dF \citep{Beutler_2011} and the SDSS DR7 MGS sample \citep{Ross:2014qpa}. The combination of SDSS$^+$+$\theta_{\star}$+DESY3+PP yields $H_0 = 68.7 \pm 1.7$~km/s/Mpc, $r_d = 145.7 \pm 3.2$ Mpc, and $\Omega_m = 0.307 \pm 0.007$, shifting the mean value of $H_0$ towards lower values. Adding the informative Gaussian prior on $A_s$ to this combination improves the constraints to $H_0 = 68.7 \pm 1.2$~km/s/Mpc, $r_d = 145.7 \pm 1.6$, and $\Omega_m = 0.307 \pm 0.007$, without altering the mean values.

With the current data, the precision of the $H_0$ constraint obtained with this method is primarily limited by how well $\Omega_m h^2$ can be determined from the lensing data. This explains the nearly identical uncertainties in $H_0$ obtained when using SDSS$^+$ or DESI2 in the absence of an informative prior on $A_s$. The $\Omega_m h^2 - H_0$ degeneracy present in the BAO data can be broken by providing information on $\Omega_m h^2$. When the uncertainty in $\Omega_m h^2$ is large, it dominates the uncertainty in $H_0$, and improvements in the quality of the BAO data do not translate in a better measurement of $H_0$. However, as the uncertainty in $\Omega_m h^2$ is reduced, there is a threshold below which the uncertainty in $H_0$ is dominated by precision of the BAO data, and the differences between BAO datasets become apparent. At present, a prior on $A_s$ is required to reach that threshold. This situation may change in the future, with increased sensitivity of weak-lensing observables to the shape of the matter power spectrum, enabling more precise determinations of $\Omega_m h^2$ without relying on external priors.

\subsection{Current data constraints with varying $\Sigma m_\nu$}

Neutrinos with masses in the $[0, 3.0]$~eV range contribute as radiation at matter-radiation equality, and also suppress the growth of cosmic structure on smaller scales. Both effects affect the shape of the matter power spectrum and, consequently, the CMB lensing convergence spectrum and the galaxy lensing and clustering, making the latter a probe of the total neutrino mass.

As discussed in Sec.~\ref{degeneracies} and seen in Fig.~\ref{fig:triangle_current_all}, there is a positive correlation between \( \Sigma m_\nu \) and \( \Omega_m h^2 \) driven by the need to preserve the value of \( k_{\rm eq} \) for larger values of \( \Sigma m_\nu \). With \( \Omega_m \) constrained by BAO (and independently also by SN), this implies that \( \Sigma m_\nu \) is positively correlated with \( H_0 \). We find that allowing \( \Sigma m_\nu \) to vary considerably dilutes the constraint on $H_0$, with the combination of Base+DESY3+PP yielding $H_0 = 75.3^{+3.3}_{-4.0}$~km/s/Mpc. Adding a CMB-informed prior on $A_s$ improves this to $H_0 = 73.9\pm 2.2$~km/s/Mpc, while constraining the neutrino masses to $\Sigma m_\nu = 0.46^{+0.21}_{-0.25}$~eV ($0.46^{+0.40}_{-0.45}$~eV at 95\% CL).

Assigning uniform probability for $\Sigma m_\nu$ to take any value in the $[0,3]$~eV range, with the parameter being relatively unconstrained positively correlated with $H_0$, tends to bias the $H_0$ and $\Omega_m h^2$ posteriors towards larger values. To demonstrate that this is a prior-related effect, in Figure~\ref{fig:log_prior_mnu} we show posteriors for a few data combinations while using a logarithmic prior on $\Sigma m_\nu$ covering the $[0.001,3]$~eV mass range. With the logarithmic prior, the combination of Base+DESY3+PP+$A_s$ yields $H_0 = 71.5^{+1.4}_{-2.9}$ and $\Sigma m_\nu = 0.205^{+0.097}_{-0.23}$ eV at 68\% CL ($0.20^{+0.46}_{-0.24}$ eV at 95\% CL). These are statistically consistent with the uniform prior results, but now without the apparent bias.

The $r_d$-agnostic bounds on $\Sigma m_\nu$ presented in this section are substantially weaker than those from the conventional derived-$r_d$ analysis of the DESI DR2+Planck+APS-L+DESY3+PP dataset, $\Sigma m_\nu < 0.022$~eV ($0.050$~eV at 95\% CL), that are getting close to being in tension with the lower limit of $0.06$~eV.

Notably, the constraints on $S_8$ derived while varying $\Sigma m_\nu$ are effectively the same as those from the fixed $\Sigma m_\nu$ analysis in the previous subsection, and in excellent agreement with $S_8=0.815^{+0.016}_{-0.021}$ obtained by the KiDS-Legacy analysis~\citep{Stolzner:2025htz,Wright:2025xka} with fixed $\Sigma m_\nu=0.06$~eV, and $S_8=0.811^{+0.022}_{-0.020}$ from the DES Year 3 joint analysis of cluster abundances, weak lensing, and galaxy clustering while varying $\Sigma m_\nu$.

\subsection{Forecasted parameter constraints}
\label{sec:forecasts}

Finally, we present a forecast based on mock CMB lensing data from a Simons-Observatory-like (SO-L) experiment,  a future $3\times2$-pt galaxy weak lensing and clustering dataset, the completed DESI BAO program, the present value of $\theta_{\star}$, a mock future SN sample with $\sim 9000$ SN magnitudes, and the same prior on $A_s$ as the one used in the earlier subsections. We also perform the same forecast with a cosmic-variance-limited (CV-L) CMB Lensing mock data. The forecasted parameter constraints for the cases of fixed and varied $\Sigma m_\nu$ are summarized in Tables ~\ref{tab:fixmnu} and \ref{tab:varymnu}, and shown in Fig.~\ref{fig:triangle_future_all}. These forecasts are not meant to be an exhaustive study of the promise of the $r_d$-agnostic method, but rather provide a sense of what improvement one can expect over a reasonable timescale.

With a fixed $\Sigma m_\nu$ and without an informative $A_s$ prior, we obtain $H_0 = 67.30^{+1.1}_{-1.2}$~km/s/Mpc from the combination including the SO-L data, and $H_0 = 67.10 \pm 1.0$~km/s/Mpc from the combination with CV-L. It is interesting that eliminating the noise from the CMB lensing dataset, as in the cosmic variance limited case, leads to almost no improvement in the constraints. This is because the uncertainty in $H_0$ is dominated by the remaining degeneracy between $\Omega_mh^2$ and $A_s$.
The inclusion of the $A_s$ prior significantly reduces the uncertainty, yielding $H_0 = 67.24 \pm 0.67$ from the combination with SO-L, and $H_0 = 67.30 \pm 0.57$ from the combination with CV-L, while also improving the constraints on other cosmological parameters. This demonstrates that the $r_d$-agnostic method is capable of constraining the Hubble constant at one-percent level with a realistic CMB Lensing dataset when $\Sigma m_\nu$ is fixed.

When allowing $\Sigma m_\nu$ to vary, we find $H_0 = 67.71^{+0.84}_{-1.1}$ km/s/Mpc and $\Sigma m_\nu <0.133 \, (<0.263)$ eV at 68\% (95\%) CL from the BAO+$\theta_{\star}$+SO-L+$3\times2$-pt+SN+$A_s$ combination, and $H_0 = 67.99^{+0.81}_{-0.97}$ km/s/Mpc and $\Sigma m_\nu = 0.123^{+0.051}_{-0.079} \, (<0.244)$ eV at 68\% (95\%) CL from BAO+$\theta_{\star}$+CV-L+$3\times2$-pt$^*$+SN+$A_s$ combination. As in the fixed $\Sigma m_\nu$ case, the added sensitivity to small-scale suppression at high $\ell$ offered by the CV-L combination does not lead to a substantial improvement in constraining power, while the prior on $A_s$ makes a notable difference.

In the mock posteriors, we no longer see the shift to higher  $\Omega_mh^2$ and $H_0$ values when sampling $\Sigma m_\nu$ with a uniform prior. This is because $\Sigma m_\nu$ is much better constrained by the mock data, reducing the dependence on the prior choice.

\section{Summary}
\label{sec:concl}

The Hubble tension, and attempts to resolve it through new ingredients in early-Universe physics, motivate efforts to measure $H_0$ without relying on a model for the sound horizon $r_d$ or calibrating supernovae. In this work, we combined CMB lensing measurements from ACT, \textit{Planck}, and SPT-3G with the DES Year-3 $3\times2$-pt galaxy weak lensing and clustering correlations, DESI DR2 BAO, the CMB acoustic scale angle $\theta_{\star}$, and PP SN data, while treating $r_d$ as a free parameter. 

With an informative prior on the amplitude of primordial fluctuations $A_s$, we find $H_0 = 70.03 \pm 0.97$~km/s/Mpc when fixing the neutrino mass sum to its minimal value $\Sigma m_\nu = 0.06$~eV. This is $2.1\sigma$ below the SH0ES Cepheid-based distance-ladder measurement~\citep{Riess:2021jrx} and $2.4\sigma$ above the \textit{Planck} $\Lambda$CDM value~\citep{Rosenberg:2022sdy}, consistent with alternative distance-ladder determinations~\citep{Freedman:2024eph}. Allowing $\Sigma m_\nu$ to vary increases the uncertainty in $H_0$, but still yields a sub-3\% $r_d$-agnostic constraint of $H_0 = 73.9 \pm 2.2$~km/s/Mpc. 

Forecasts combining future CMB lensing data from a Simons-Observatory-like experiment, a next-generation $3\times2$-pt dataset, the completed DESI BAO program, $\theta_{\star}$, a future SN sample, and an informative $A_s$ prior predict $\sigma(H_0) \simeq 0.67$~km/s/Mpc. This is comparable to the precision of full-\textit{Planck} CMB analyses, demonstrating that the sound-horizon-agnostic approach can deliver an independent and competitive test of the need for new physics at recombination.

While the informative $A_s$ prior used in this work encompasses the parameter space allowed by modified recombination models with higher $H_0$, it will be valuable to explore whether future tomographic clustering and weak-lensing data from Vera Rubin LSST~\citep{LSSTDarkEnergyScience:2018jkl} and \textit{Euclid}~\citep{Amendola:2016saw} can remove the need for such a prior. The expanded mode coverage from LSST and \textit{Euclid} should improve sensitivity to the neutrino free-streaming suppression of growth~\citep{Mishra-Sharma:2018ykh} and help disentangle the effects of $\Omega_m h^2$, $A_s$, and $\Sigma m_\nu$, especially when combined with low-noise CMB lensing. We leave this for future work.

\begin{acknowledgments}
We thank K. N. Abazajian, D. Alonso, K. Jedamzik, L. Legrand, J. Lesgourgues, S. Raghunathan, R. K. Sharma, B. D. Sherwin and G. B. Zhao for helpful discussions. This research was enabled in part by support provided by the BC DRI Group and the Digital Research Alliance of Canada ({\tt alliancecan.ca}). SHM and LP are supported by the National Sciences and Engineering Research Council (NSERC) of Canada.
\end{acknowledgments}

\FloatBarrier

\bibliographystyle{aasjournalv7}

\end{document}